\documentclass[fleqn,usenatbib]{mnras}

\usepackage{newtxtext,newtxmath}
\usepackage{appendix}
\usepackage{makecell}
\usepackage{caption}
\usepackage{float}
\usepackage[T1]{fontenc}
\usepackage{multirow} 
\DeclareRobustCommand{\VAN}[3]{#2}
\let\VANthebibliography\thebibliography
\def\thebibliography{\DeclareRobustCommand{\VAN}[3]{##3}\VANthebibliography}

\usepackage{subfigure}
\usepackage{placeins}
\usepackage{soul}
\usepackage{graphicx}	
\usepackage{amsmath}	
\usepackage[dvipsnames]{xcolor}

\usepackage{comment}
\usepackage{array}

\usepackage{xltabular}
\usepackage{pdflscape}

\title[UVIT view of NGC5291]{UVIT view of NGC 5291: Ongoing star formation in tidal dwarf galaxies at $\sim$ 0.35 kpc resolution} 

\author[Rakhi et al]{Rakhi R $^{1}$\thanks{E-mail: rakhi@nsscollegepandalam.ac.in},
Geethika Santhosh $^{1}$,
Prajwel Joseph $^{2,3}$,
Koshy George $^{4}$,
Smitha Subramanian $^{2}$,
\newauthor Indulekha Kavila $^{5}$,
 J. Postma $^{6}$,
Pierre-Alain Duc $^{7}$,
Patrick C{\^o}t{\'e} $^{8}$,
Luca Cortese  $^{9}$,
S. K. Ghosh  $^{10}$,
\newauthor Annapurni Subramaniam $^{2}$,
Shyam Tandon $^{11}$,
John Hutchings $^{8}$,
P Samuel Wesley $^{3}$,
Aditya Bharadwaj $^{3}$,
\newauthor Neeran Niroula $^{3}$
\\\\
$^{1}$ Department of Physics, NSS College, Pandalam, Kerala 689 501, India  \\
$^{2}$ Indian Institute of Astrophysics, Bangalore 560034, India\\
$^{3}$ Department of Physics and Electronics, CHRIST (Deemed to be University), Bangalore 560029, India\\
$^{4}$ Faculty of Physics, Ludwig-Maximilians-Universit{\"a}t, Scheinerstr. 1, 81679, Munich, Germany\\
$^{5}$ 	School of Pure and Applied Physics, Mahatma Gandhi University, Kottayam, Kerala 686560 India\\
$^{6}$ University of Calgary, 2500 University Drive NW, Calgary, Alberta, Canada \\
$^{7}$ Universit{\'e} de Strasbourg, CNRS, Observatoire astronomique de Strasbourg, UMR 7550, F-67000 Strasbourg, France\\
$^{8}$ Herzberg Astronomy and Astrophysics Research Centre, National Research Council of Canada, 5071 W. Saanich Road, Victoria, BC, V9E 2E7, Canada\\
$^{9}$ International Centre for Radio Astronomy Research (ICRAR), University of Western Australia, Crawley, WA 6009, Australia\\
$^{10}$ Tata Institute of Fundamental Research, Colaba, Mumbai 400005, India\\
$^{11}$ Inter-University Centre for Astronomy and Astrophysics, PostBag 4, Ganeshkhind, Pune-411007, India\\
}

\pubyear{2021}

\begin{document}

\label{firstpage}
\pagerange{\pageref{firstpage}--\pageref{lastpage}}
\maketitle

\begin{abstract}

NGC 5291, an early-type galaxy surrounded by a giant HI ring, is believed to be formed from collision with another galaxy. Several star forming complexes and tidal dwarf galaxies are distributed along the collisional ring which are sites of star formation in environments where extreme dynamical effects are involved. Dynamical effects can affect the star formation properties and the spatial distribution of star forming complexes along the tidal features. To study and quantify the star formation activity in the main body and in the ring structure of the NGC 5291 system, we use high spatial resolution FUV and NUV imaging observations from the Ultraviolet Imaging Telescope onboard AstroSat.  A total of 57 star-forming knots are identified to be part of this interacting system out of which 12 are new detections (star forming complexes that lie inside the HI contour) compared to the previous measurements from lower resolution UV imaging. We estimate the attenuation in UV for each of the resolved star-forming knots using the UV spectral slope $\beta$, derived from the $FUV-NUV$ colour. Using the extinction corrected UV fluxes, we derive the star formation rate of  the resolved star forming complexes. The extinction corrected total star formation rate of this system is estimated as 1.75 $\pm$ 0.04  $M_\odot/yr$.  The comparison with dwarf galaxy populations (BCD, Sm and dIm galaxies) in the nearby Universe shows that many of the knots in the NGC 5291 system have SFR values comparable to the SFR of BCD galaxies. 

\end{abstract}


\begin{keywords}
galaxies: star formation – galaxies: interactions – galaxies: dwarf – galaxies: formation – ultraviolet: galaxies-stars: formation
\end{keywords}


\section{Introduction}
Study of galaxy mergers and interactions are of great importance in advancing our current understanding of 
 galaxy formation and evolution 
 \citep{2003AJ....126.1183C, 2019A&A...631A..51P}.
In the hierarchical $\Lambda$ Cold Dark Matter ($\Lambda$CDM) clustering paradigm, large massive 
 halos form and grow by the merging or clustering of 
low mass halos. 
The hierarchical clustering model predicts that massive galaxies must have undergone several merging activities in the past. This leads to the possibility of dwarf galaxies being the primary ingredients in the formation of the large galaxies we see in the nearby universe
\citep{1978MNRAS.183..341W,1991ApJ...379...52W}.

The local Universe is observed to have several interacting or merging galaxy systems characterized by dust and gas-rich tidal tails, collisional rings and tidal bridges \citep{1993AJ....106..939B, 1995ApJ...455..524H, 1998AstL...24...73S}.
 Tidal dwarf galaxies (TDGs) are gravitationally bound systems of gas and stars formed during interaction of galaxies and are kinematically decoupled from the surrounding tidal debris.
During close encounters between gas-rich galaxies, neutral hydrogen gas (HI), stars and dust from the disks of the galaxies can get pulled out by tidal forces/gravitational torques \citep{Bournaud_2010} forming rings, tidal tails, bridges and plumes. 
star formation takes place in the gas thrown out of the galaxies during tidal interactions.  
Star forming knots or clumps are observed along the tidal features \citep{1999PhR...321....1S} and the 
massive clumps are potential young TDG candidates 
\citep{1992A&A...256L..19M, 1993ApJ...412...90E, 2000ApJ...532..845A,  2004A&A...427..803D, 2009AJ....137.4643H, 2012ASSP...28..305D}.
The most massive TDGs in an interacting system may evolve to become self-bound dwarf galaxies that may detach from the host system \citep{1999IAUS..186...61D}.  Once separated from their progenitors, they will closely resemble the independent dwarf galaxy populations. Being
pre-enriched these TDGs are more metal rich than isolated dwarf galaxies of the same luminosity. This property of TDGs can be used to identify these recycled dwarf galaxies and to investigate the origin of their building material in the disk of their progenitors \citep{2000ApJ...542..137H}. 

TDGs comprise of young stars, which are formed from  the recent collapse of ejected HI clouds as well as the older stellar population coming from the disk of their parent galaxies. \cite{1999IAUS..186...61D} studied the relative proportion of both these populations using multi-wavelength observations of several interacting systems in the nearby Universe. They proposed that TDGs are divided into two categories. Category 1 consists of extremely young objects, forming their first generation of stars (e.g. dwarfs around NGC 5291). These have high star formation rates (SFR) similar to that of blue compact dwarf (BCD) galaxies.  Category 2 corresponds to galaxies dominated by the older stellar population coming from the disk of their progenitors and these galaxies resemble dwarf irregulars (e.g. NGC 2992)  \citep{ 2000AJ....120.1238D, Bournaud_2010}.

Young, hot, massive and luminous O, B, A stars on the main sequence give out immense amount of ultraviolet (UV) radiation and therefore regions of ongoing star formation could appear bright in ultraviolet images. The ultraviolet continuum is thus a direct tracer of recent star formation in galaxies  ($\sim$ 200 Myr) \citep{2012ARA&A..50..531K}. With the advent of UV missions capable of providing deep and high resolution
UV images of extragalactic systems, a quantitative analysis of the star formation activity in  star forming knots in terms of the SFR is possible. Tidal dwarf galaxy formation is connected to merging or interacting galaxies in the universe \citep{2000ApJ...543..149O}. TDGs with ongoing star formation 
are important structures to study the process of star formation in the smallest mass systems (dwarf galaxies) also.
NGC 5291 is an interacting galaxy system that lies in the western outskirts of the cluster Abell 3574. 
The system comprises of an early-type galaxy NGC 5291 (morphological type: SA.0+ ) and a companion galaxy called "the Seashell" (morphology: distorted edge-on spiral) interacting with it \citep{1979MNRAS.188..285L}. 

The system has extensions or tails, defined by knots, emerging from the galaxy. Deep optical and spectroscopic studies of NGC 5291 pointed out that the optical knots that extend to the north and south of the system may be sites of recent star formation \citep{1978Msngr..13...11P, 1979MNRAS.188..285L}.
21 cm radio observations, using the Very Large Array (VLA), revealed a giant collisional HI ring structure connected to the NGC 5291 system which indicated that the knots observed are indeed star forming complexes that may even be young tidal dwarf galaxies 
\citep{1997AJ....114.1427M, 2007Sci...316.1166B}.
The fragmented HI ring structure, which hosts numerous intergalactic HII regions of the NGC 5291 system, is exceptional in itself because of its moderately high metallicity $\left(8.4\le 12 + log (O/H) \le 8.6\right)$ and the absence of an old stellar population. This suggests that the dwarf galaxies observed in the system are 
in fact young tidal dwarf galaxies formed from the pre-enriched gas in the collisional/tidal debris \citep{1998A&A...333..813D}.  
\\

Many studies of the NGC 5291 system in the ultraviolet have come up over the past few decades consequent to
observations using the Far Ultraviolet Space Telescope (FAUST) \citep{1984Sci...225..184B}, the Hubble Space Telescope (HST) and the Galaxy Evolution Explorer (GALEX)  \citep{1994A&A...289..715D,2007A&A...467...93B,2009AJ....137.4561B,2019yCat..36280060F,2020ApJ...888L..27E}. Among these, GALEX is fully dedicated to observations in the ultraviolet regime and is capable of providing wide field (1.2\textdegree) far ultraviolet (FUV) and near ultraviolet (NUV) images  with a spatial resolution of 4.2$\arcsec$/5.3$\arcsec$ (FUV/NUV) \citep{2007ApJS..173..682M}. \citet{2007A&A...467...93B} presented a polychromatic view of NGC 5291 based on GALEX observations together with archival H$\alpha$, 8 $\mu$m and HI data. They identified 29 star forming regions along the ring structure and determined their SFR.
More recently,  \citet{2019yCat..36280060F}, using HST data, studied massive star cluster formation in the three TDGs (NGC 5291N, NGC 5291S and NGC 5291 SW)  associated with the NGC 5291 system.\\ 

\begin{figure*}
 \includegraphics[scale=1.0] {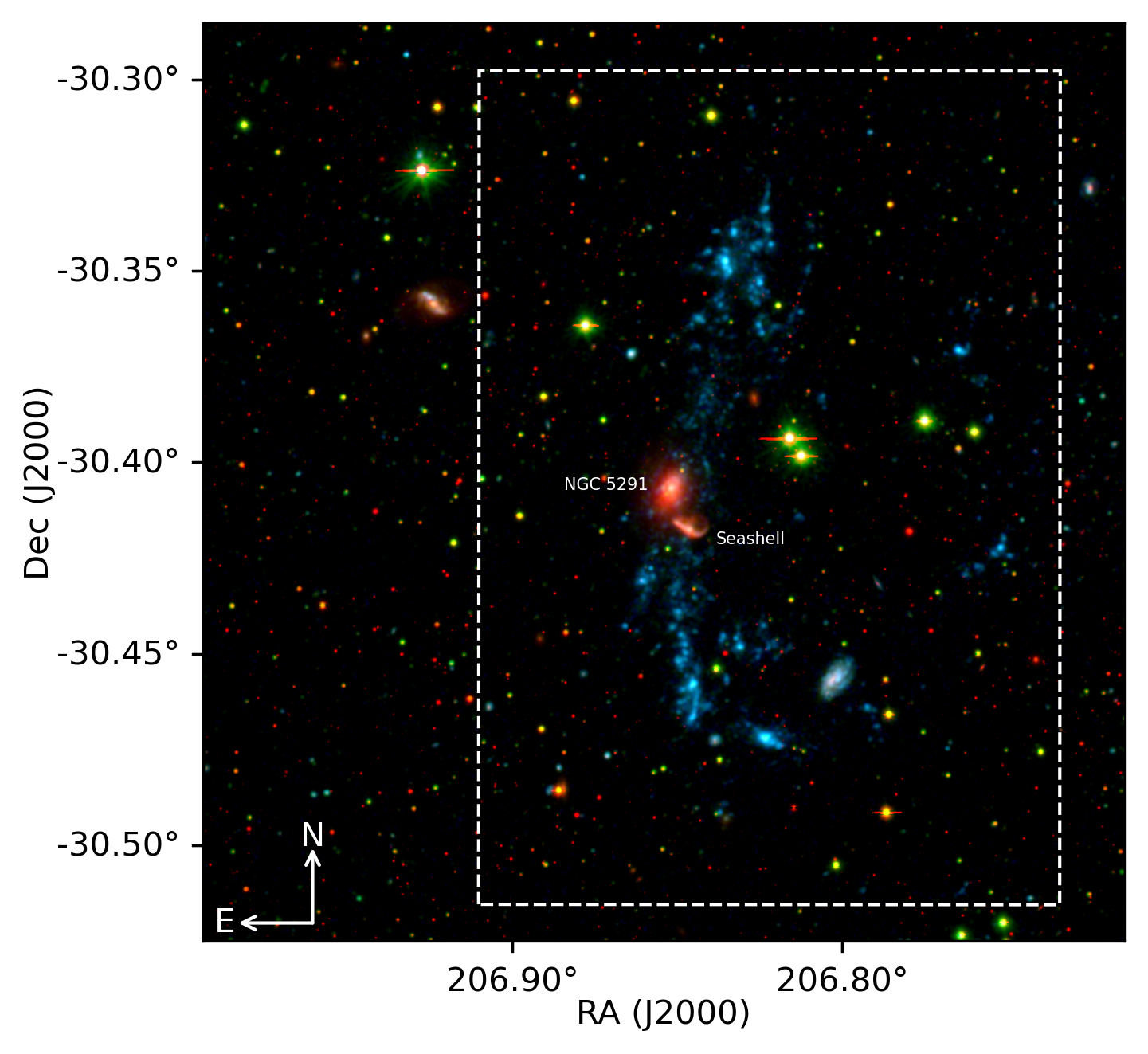}
 \caption{colour composite image of the NGC 5291 system made using FUV (blue), NUV (green) and DECaLS z band image (red). The dashed rectangle shows the region of interest.
 (Field of View: 21.8 $\arcmin$ $\times$ 11.6 $\arcmin$)}
 \label{fig:rgb image ngc5291}
\end{figure*}

In this paper,  we present high resolution ultraviolet imaging observations (in Far and Near UV bands, FUV: 1.4$\arcsec$ and NUV: 1.2$\arcsec$) of the NGC 5291 system using 
data from the Ultraviolet Imaging Telescope (UVIT) on board AstroSat. 
The main aim of the paper is to identify and characterize the star forming knots in the tidal tails and determine the star formation rates in these knots at the best possible resolution, taking into account dust attenuation of the ultraviolet spectrum.
This paper is outlined as follows. Section 2 of the paper describes data acquisition, data reduction, source extraction and identification in detail. The results are discussed in Section 3 and conclusions are presented in Section 4.\\

Throughout the paper, magnitudes are computed in AB system \citep{1983ApJ...266..713O}. 
The value of the Hubble parameter $H_0$ used is 72 km s$^{-1}$ Mpc$^{-1}$, assuming flat $\Lambda$CDM cosmology. For this value of $H_0$, the distance (D) to NGC 5291 is 62 Mpc \citep{2007A&A...467...93B} and 1" in the sky corresponds to 0.296 kpc at system rest frame.
\section{Data \& Analysis}
\begin{table}
 \caption{\label{t1} NGC 5291 UVIT observations} 
 \label{galaxy details}
 \tabcolsep=0.3cm
 \begin{tabular}{ccccc} 
 \hline 
 Channel & Filter Name  & 
 \makecell{$\lambda$\textsubscript{mean} \\ ({\AA})} &  \makecell{$\Delta \lambda$ \\ ({\AA})}  & \makecell{Integration time \\ (s)} \\
 \hline
 FUV & F148W & 1481 & 500 & 8242 \\
 NUV & N242W & 2418 & 785 & 8079 \\
 \hline
 \end{tabular}
 \end{table}

NGC 5291 (RA:206.852,Dec:-30.407)\footnote{https://ned.ipac.caltech.edu/} has been observed (PI: K. George, proposal ID: G07\_003) with Ultraviolet Imaging Telescope (UVIT) on board 
AstroSat 
\citep{2012SPIE.8443E..1NK}.
UVIT  performs imaging simultaneously in three channels:  visible (320-550 nm), the near-ultraviolet (NUV: 200-300 nm) and the far-ultraviolet (FUV : 130-180 nm). UVIT has a set of filters mounted on a wheel to facilitate imaging in the NUV and FUV 
in different narrow and broad wavelength bands. The field of view of UVIT is 28$\arcmin$ in diameter. UVIT has a resolution of $\sim$1.4$\arcsec$ in FUV and $\sim$ 1.2$\arcsec$ in NUV. This implies that UVIT can resolve star forming knots in NGC 5291  down to  approximately 0.35 kpc at NUV and 0.41 kpc at FUV.

\subsection{Data}

We use 
Level 1 (L1) UVIT data of
NGC 5291. 
For the multiple orbit observations of the target field NGC 5291, the filters used are NUV: N242W  and FUV: F148W. Details on the UVIT filter combinations  and the performance parameters for the individual filters are given in \cite{2017AJ....154..128T}. We have reported the filter details and the integration time of the UVIT observations of NGC 5291 in Table \ref{t1}. 
A HI map of the galaxy, obtained with the  VLA  \citep{2007Sci...316.1166B}  is used for identifying the relevant star-forming knots. To check whether the detections are bonafide knots, we use the observations from the Dark Energy Camera Legacy Survey (DECaLS) DR10 imaging data in three optical filters (g, r, z) \citep{2019AJ....157..168D} \footnote{https://www.legacysurvey.
org/viewer}.
\subsection{Data Reduction}
L1 data of NGC 5291  is  reduced to Level 2 (L2) scientific images using  
CCDLAB \citep{2017PASP..129k5002P,2021JApA...42...30P}. 
Using CCDLAB, UVIT
 data is corrected for fixed pattern noise, distortion and drift, and flat fielded. The orbit-wise images are aligned to a common frame before merging the 
data. The PSF of master NUV and FUV images are optimized. Finally, the images are aligned with respect to the sky coordinates using the automated WCS solver in CCDLAB \citep{2020PASP..132e4503P}. 
The NUV and FUV images have 4096 $\times$ 4096 pixel
array size where one pixel corresponds to 0.416$\arcsec$.

The NUV and FUV images thus created are used for further analysis. Flux calibration is done for NUV and FUV images using the zero point and unit conversion factors given in \citet{2017AJ....154..128T} and updated in \citet{2020AJ....159..158T}. 
\section{NGC 5291 UV imaging}
Fig. \ref{fig:rgb image ngc5291} shows the false-colour combined image of the NGC 5291 system created from the UVIT FUV and NUV images (North is up and East is towards the left of the image) overlaid with Legacy survey z band image . Here, FUV is given in blue, NUV in green  and DECaLS z band image is given in red. The interacting galaxies NGC 5291 and the Seashell are located towards the center of the image.  As seen in Fig. \ref{fig:rgb image ngc5291},  several UV bright knots extend towards the north, south and south-west directions following the fragmented ring structure seen in HI imaging data. 

\subsection{Source extraction}

The sources from the FUV and NUV images are extracted using the photometry package ProFound \citep{2018MNRAS.476.3137R}. ProFound is capable of both source identification and photometric extraction. It detects sources in noisy images, then generates segmentation maps 
by identifying the pixels belonging to each source, and measures statistics including flux, size, and ellipticity. ProFound first detects pixels from the intensity map that are above a threshold value and these pixels are allowed to grow or dilate freely until a certain intensity limit is reached based on the set threshold. The dynamic dilation will ensure close to total magnitudes regardless of the differing PSFs. 
ProFound uses watershed algorithm for de-blending pixels that are above the threshold value. The deblended collection of pixels that are above a threshold is called a segment.  This method of segmentation is unique as it does not assume a fixed aperture for an object and the segment corresponding to a source agrees with the underlying morphology of the source. After segmentation, ProFound extracts the image data flux from each of the pixels that are part of the segments. 

ProFound was run on the slightly lower resolution FUV image. Along with the FUV image, the sky and sky RMS values were given as inputs to the ProFound function and it extracted 206 sources along with their flux,
magnitude and area for sources from the entire UVIT field of view. For NUV source extraction, we provided the dilated FUV segmentation map created by ProFound as an additional input to the function to make sure that the same source position is used for both NUV and FUV. 

\begin{figure}
    \includegraphics[scale=0.14]{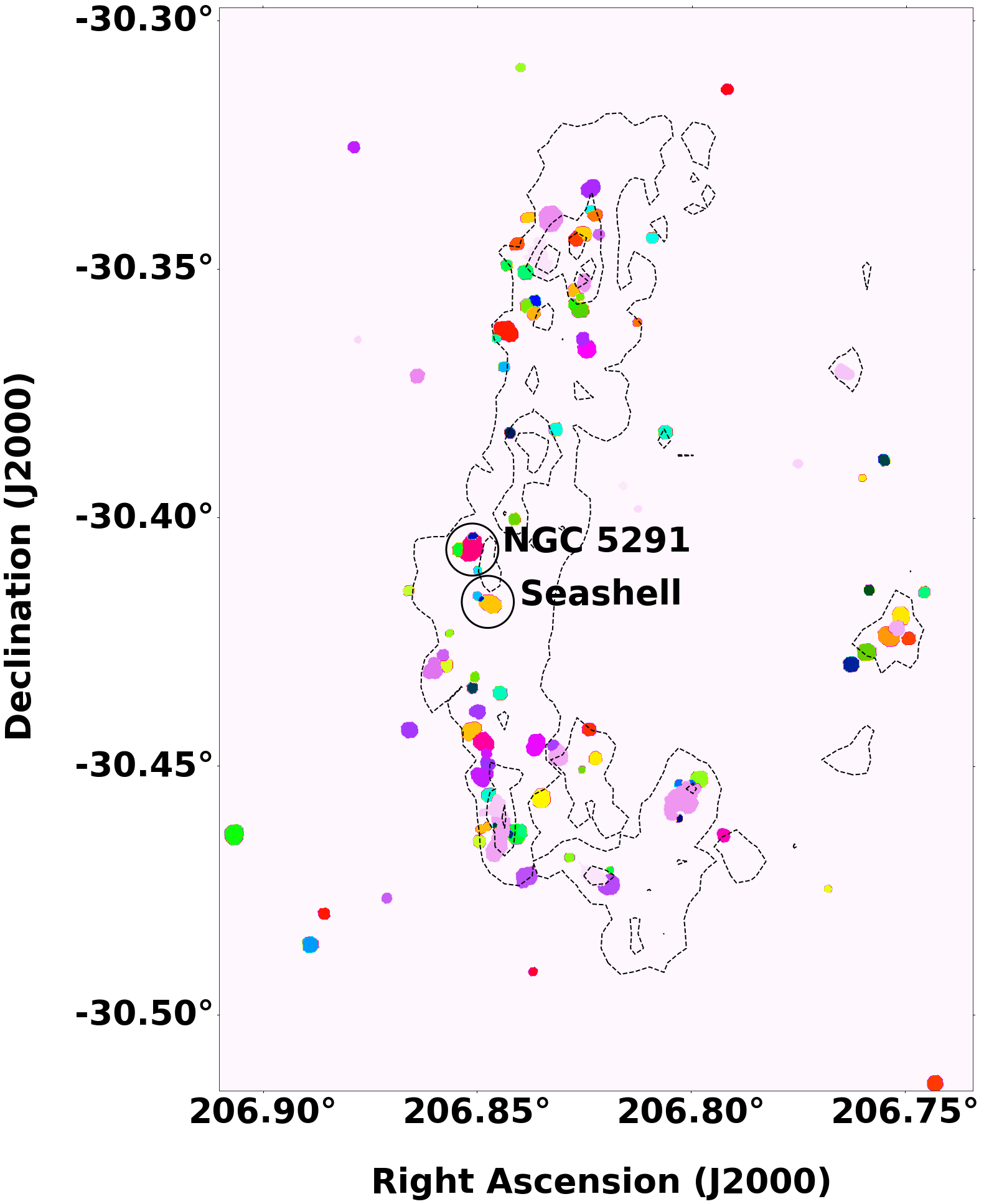}
    \caption{Segmentation map from ProFound overlaid by HI contour. The circled regions correspond to the galaxies, NGC 5291 and Seashell.}
    \label{fig:segmap}
\end{figure}

 The segmentation map obtained from ProFound, overlaid by HI contour, showing the extend of each UV source is given in Fig. \ref{fig:segmap}. ProFound assigns a unique number called the segID for each segment of the map. 
 Different colors in Fig. \ref{fig:segmap} corresponds to different segIDs. The circled regions correspond to the galaxies, NGC 5291 and Seashell.
 \subsection{Identification of star-forming knots}

\begin{figure}
 \includegraphics[scale=0.13]{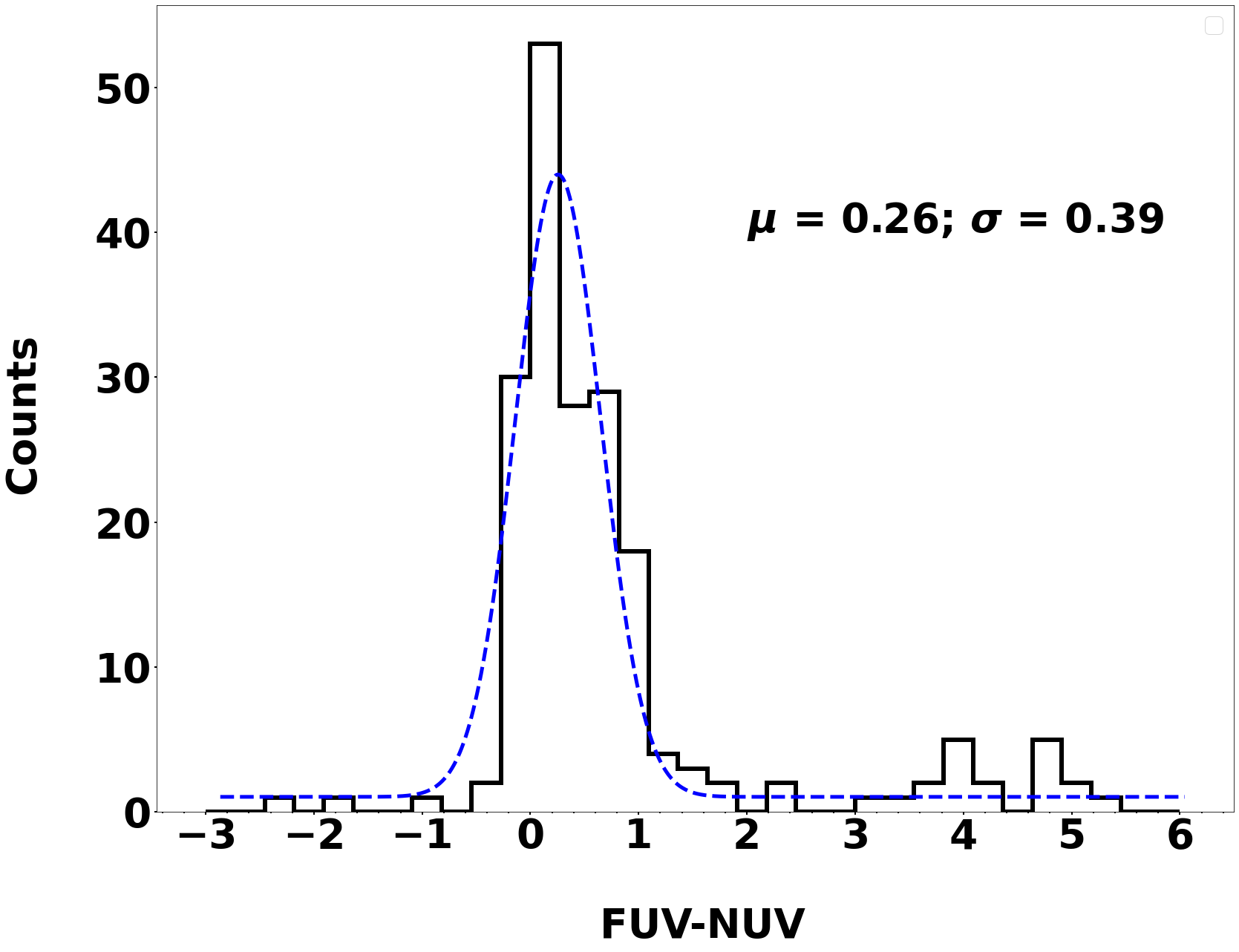}
 \caption{Distribution of FUV-NUV colour  of all the sources above 5$\sigma$ detection threshold in both FUV and NUV}
 \label{fig:fuv-nuv image ngc5291}
\end{figure}

In order to identify the star-forming knots that are part of the NGC 5291 interacting system, we analysed the FUV-NUV colour distribution of all the sources in the UVIT field of view  with a signal to noise ratio (SNR) greater than 5. 
The Gaussian fitted histogram of the UV colour (FUV-NUV) of these sources is shown in Fig. \ref{fig:fuv-nuv image ngc5291}. The mean ($\mu$), standard deviation ($\sigma$), and FWHM of the best fit are 0.26 mag, 0.39 mag, and 0.93 mag respectively.
From this, we consider those 109 sources that lie within a 1$\sigma$ colour  range, for further analysis. 
Due to the lack of redshift information for the knots, the sources that are part of the system are identified based on location within the HI contour. 
 
 \subsubsection{Sources lying within the HI contour}
All the knots that fall within the HI contour are identified with the help of SAOImageDS9 \citep{2003ASPC..295..489J}.
 Out of the 109 sources that lie within  1$\sigma$ colour  range, 64 sources 
 lie inside the HI column density contour and 45 sources lie outside (Fig. \ref{fig:segmap}). 
 \begin{figure*}
    \centering
     \subfigure[]{\includegraphics[scale=0.13]{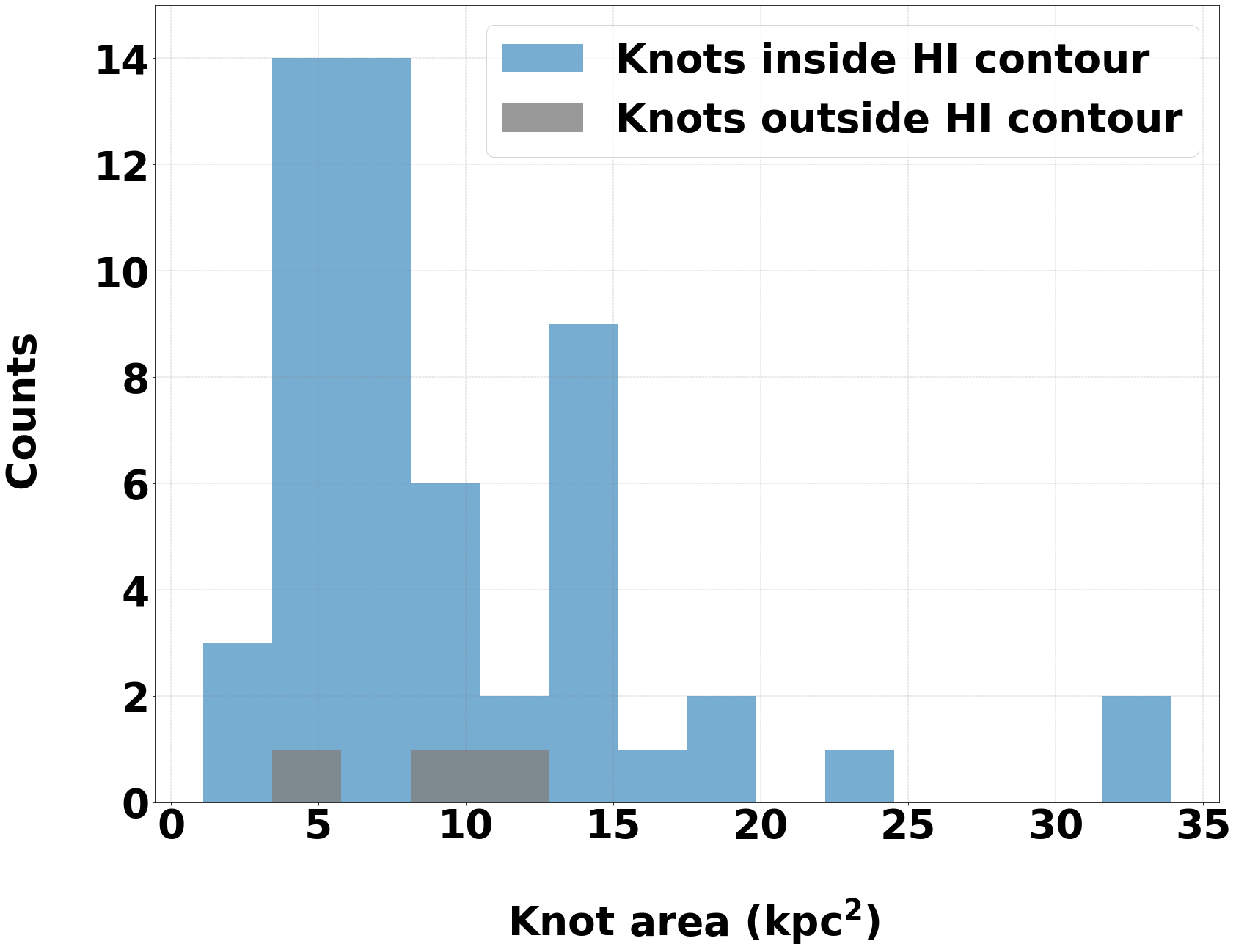}}
    \subfigure[]{ \includegraphics[scale=0.13]{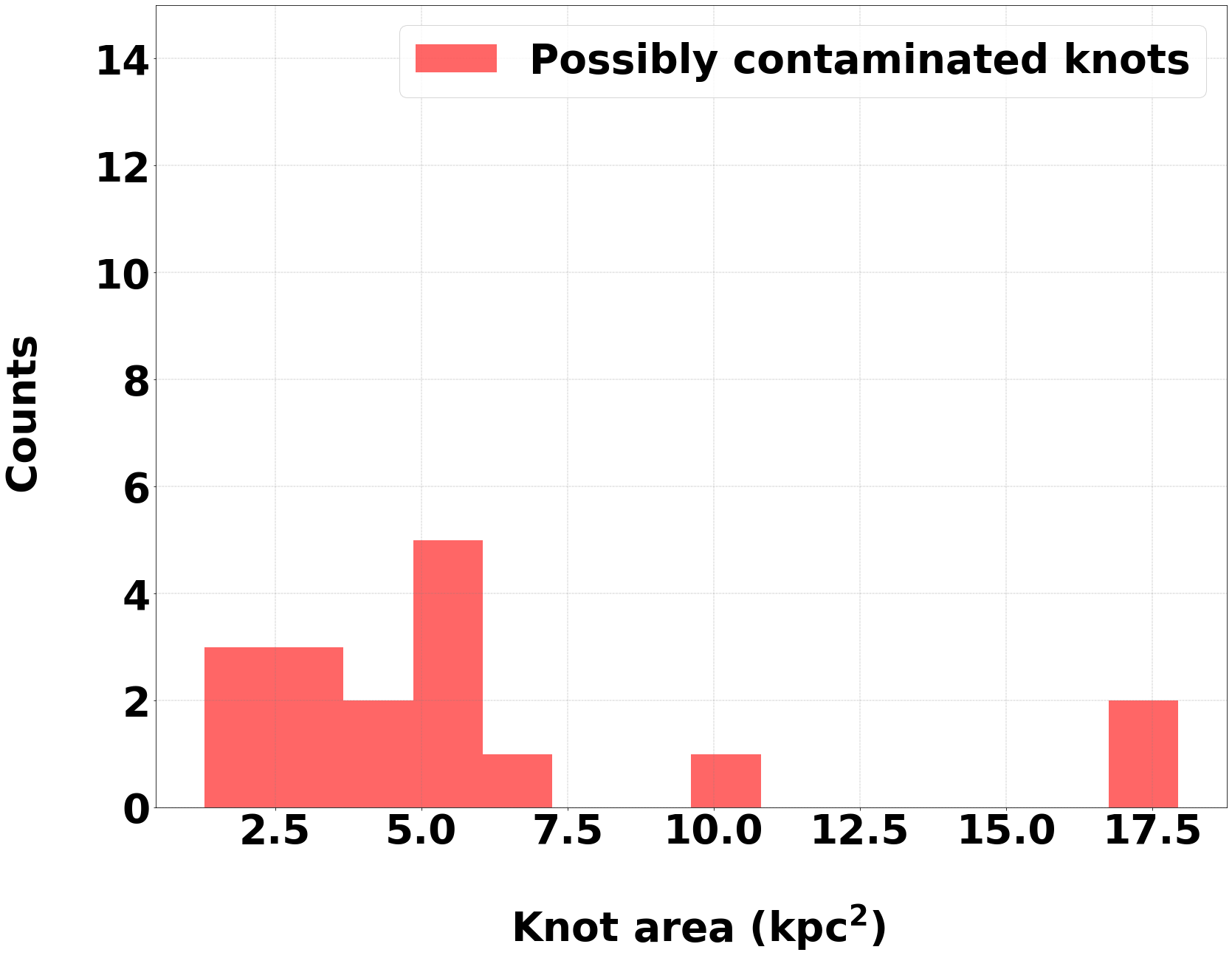}}
    \quad
    \caption{Distribution of areas (in kpc$^2$) of the resolved knots: (a) Uncontaminated regions  (b) Possibly contaminated regions  }
    \label{fig:knot and area}
\end{figure*}

 \subsubsection{Sources lying outside the HI contour}
Among the 45 sources that lie outside the HI contour,  some lie far away from the NGC 5291 interacting system. Hence we consider only those knots which are lying nearer to the HI contour for further analysis. For this, we consider the sources that are lying within a circle, which is centered at the NGC 5291-central galaxy and has a radius roughly half the diameter of the projected HI ring.  We find that a total of 10 sources out of 45 lie within this region.

In order to confirm that the selected sources (both inside as well as outside the HI contour) are star-forming regions and to check for any possible contamination, we make a comparison of these knots with  DECaLS image of the system. 
  The star-forming knots appear blue in the DECaLS image and other knots seem to be contaminated by foreground/background sources.  Thus we further eliminate a total of 17 sources (10 inside the HI contour and 7 outside the HI contour) which are possibly foreground/background sources
  We finally have a total of  57 \textit{star-forming (SF) knots} (54 inside the HI contour and 3 outside the HI contour) for the present study.
 
\begin{figure*}
\centering
\includegraphics[scale=0.10]{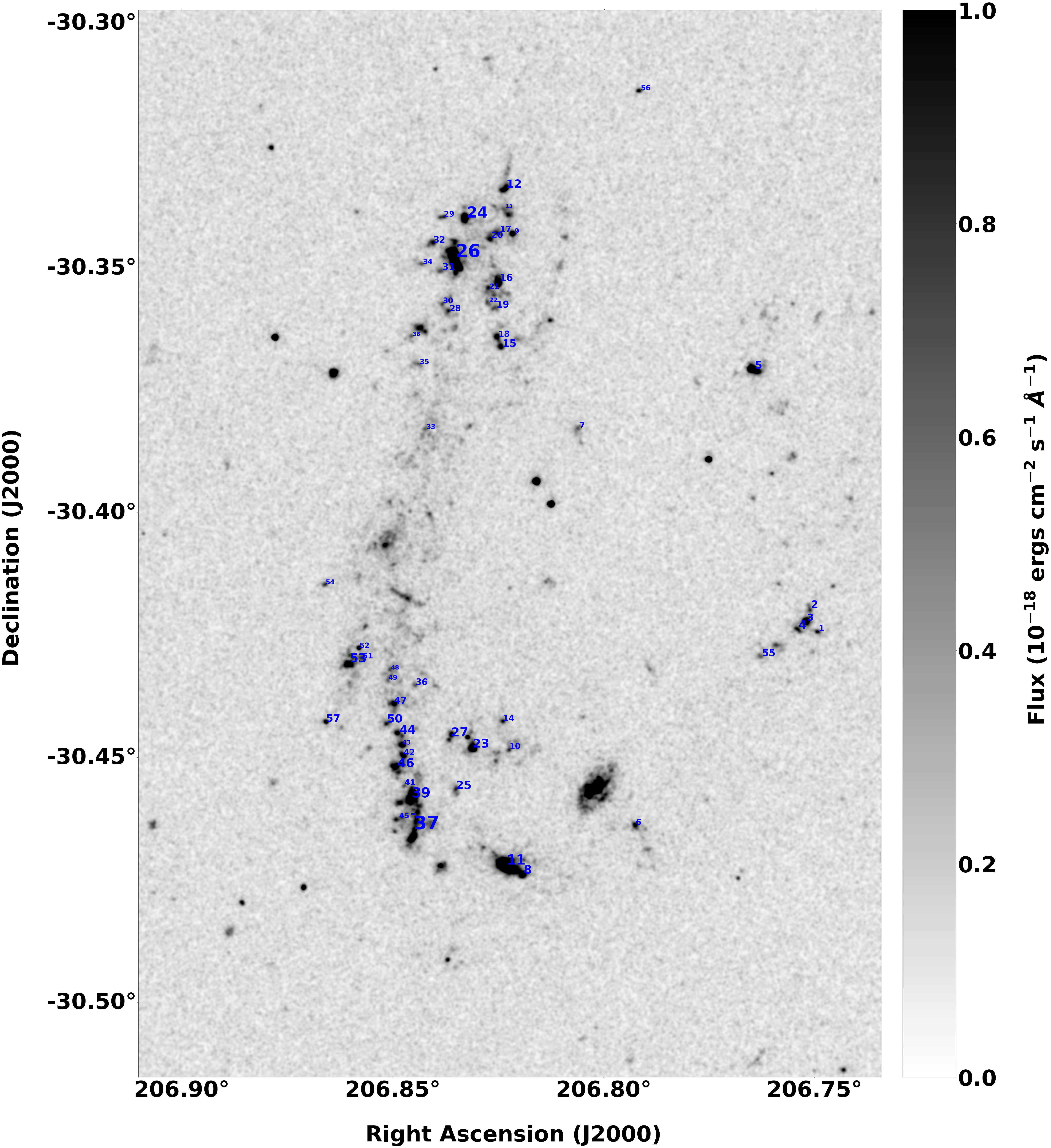}
   \caption{Selected star forming knots of the NGC 5291 system marked on FUV image. 
    The individual knots are labeled using numbers, and the sizes of these labels represent the relative sizes of the knots. Annotated regions 1 to 57 are the knots.}
    \label{fig:knots}
\end{figure*}

Fig. \ref{fig:knot and area} depicts the distribution of UV clump sizes for the resolved knots. The selected star forming knots of the NGC 5291 system (using FUV) 
are shown Fig. \ref{fig:knots}.


 \clearpage

\clearpage
\section{Results}

\subsection{Comparison with previous UV observations}

\begin{figure}
\centering
\includegraphics[scale=0.13]{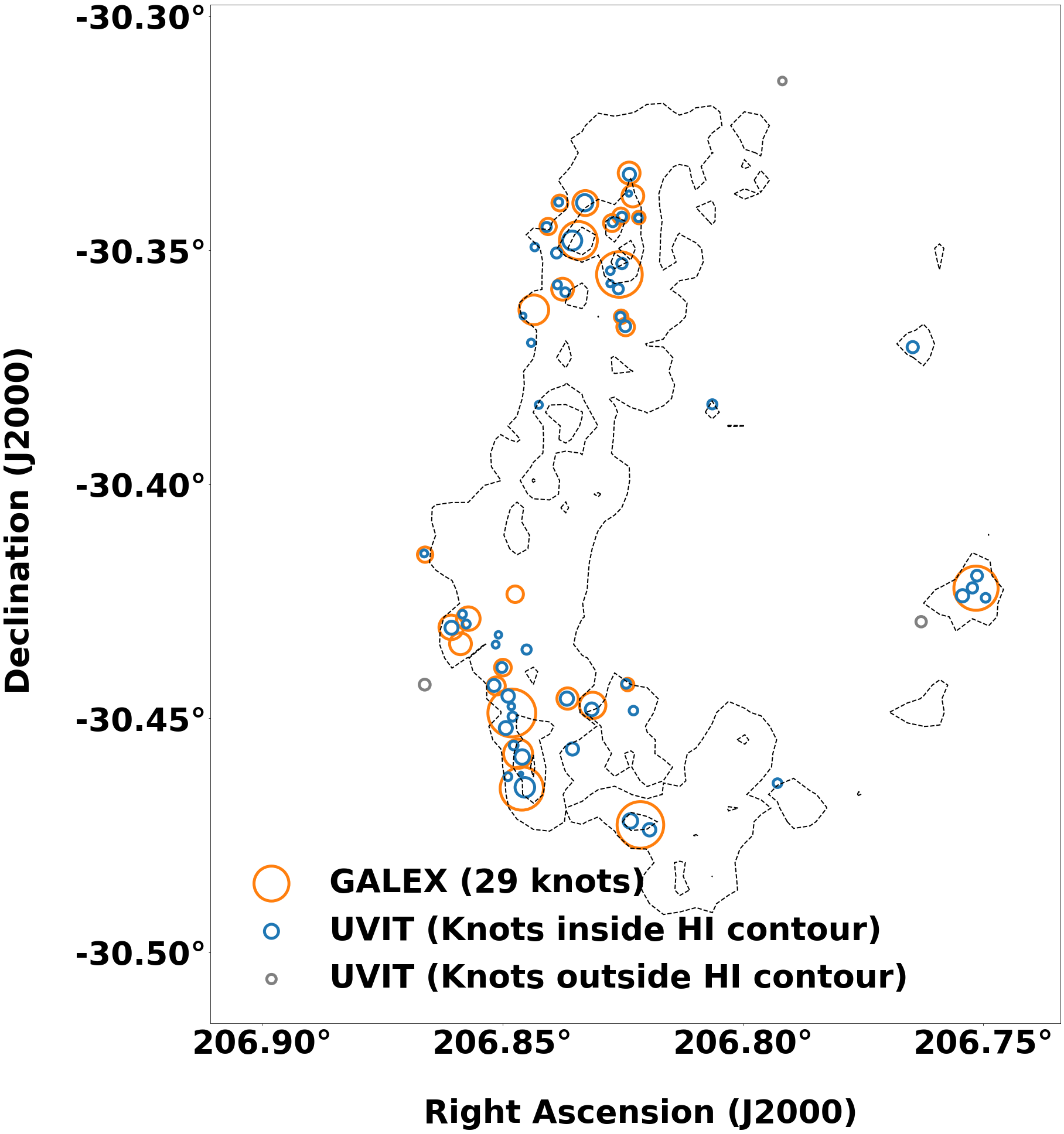}
\caption{Knots in the GALEX study \citep{2007A&A...467...93B,2009AJ....137.4561B} overlaid on the scatter plot of selected knots in the present study. The sizes of the markers for the knots represent their relative sizes.} 
\label{fig:uvit-galex}
\end{figure}

\begin{figure}
    \centering
    \subfigure[UVIT]{\includegraphics[scale=0.13]{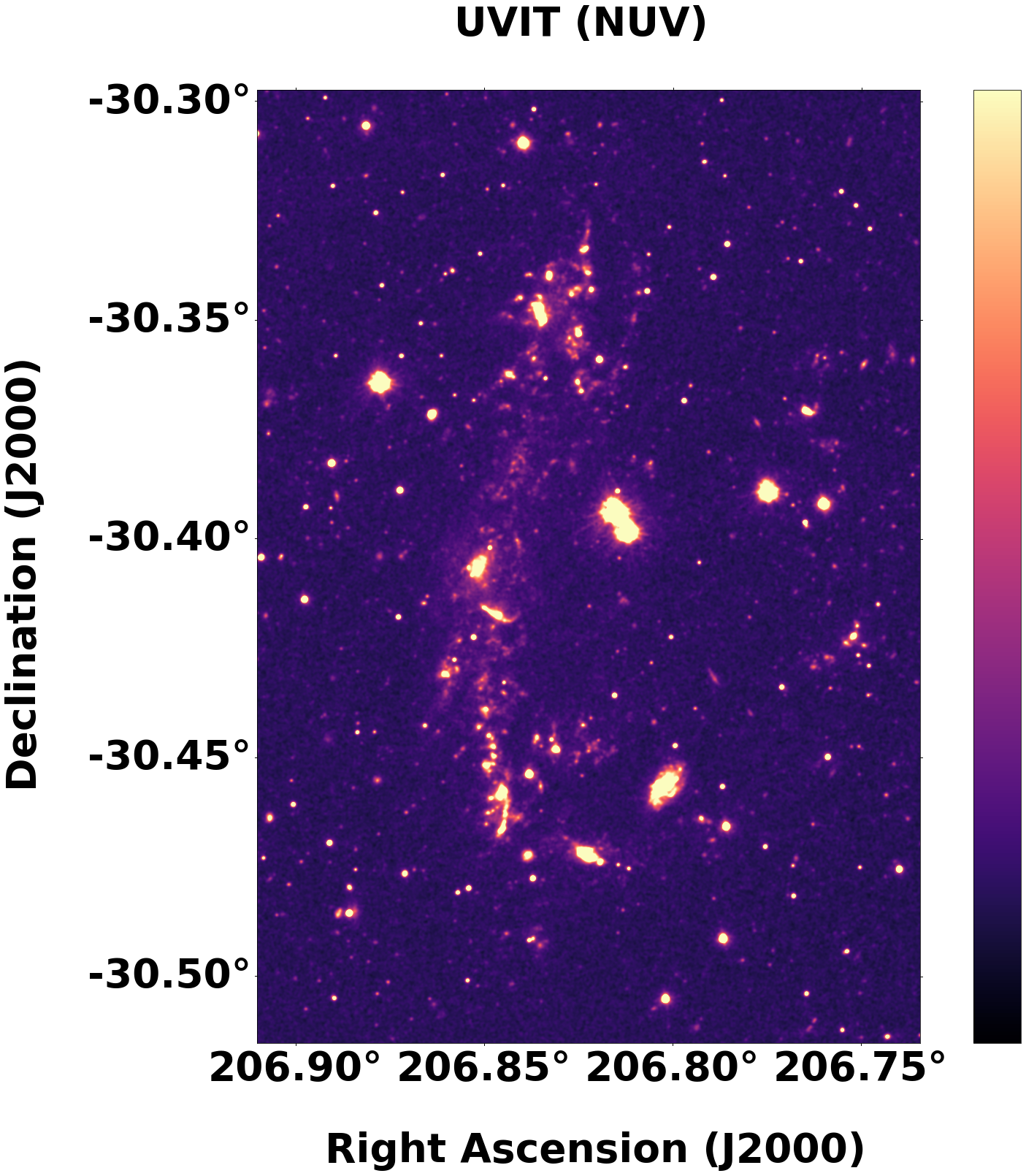}}\quad
    \hspace{3em}
    \subfigure[GALEX]{\includegraphics[scale=0.13]{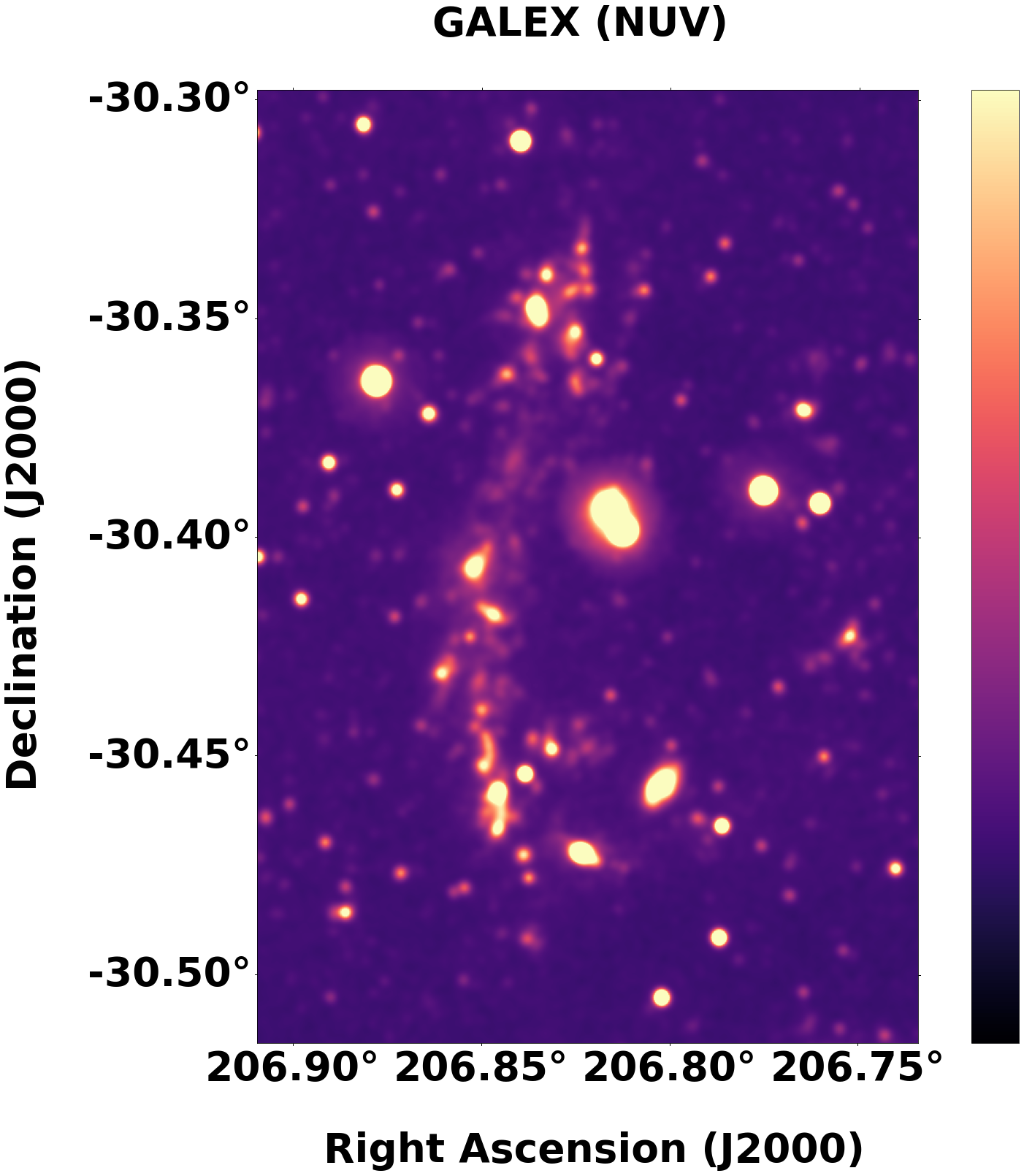}}
    \caption{Comparison of NUV images for NGC 5291 taken with UVIT (top) and GALEX (bottom). North is up and east is to the left.}
    \label{fig:galex}
\end{figure}

We first compare our high resolution UV image 
with the results from the earlier UV mission GALEX, which has a resolution $\sim$ 4-5 arcsec. The 54 star forming regions of the NGC 5291 system that lie inside the HI contour  obtained with UVIT
 have been compared against GALEX. The study of the NGC 5291 system using GALEX  reported 29 knots within the same region \citep{2007A&A...467...93B,2009AJ....137.4561B}. A comparison of the selected knots in the present study to the knots in the GALEX study is shown in Fig. \ref{fig:uvit-galex}. 
We note that 12 of the 54 UVIT knots selected within the HI contour are unreported in the GALEX-based study. It is further observed that several of the knots which appeared as a single entity in the GALEX images are well resolved by UVIT into two or more knots (see Fig. \ref{fig:galex}).

\begin{figure}
\includegraphics[scale=0.13]{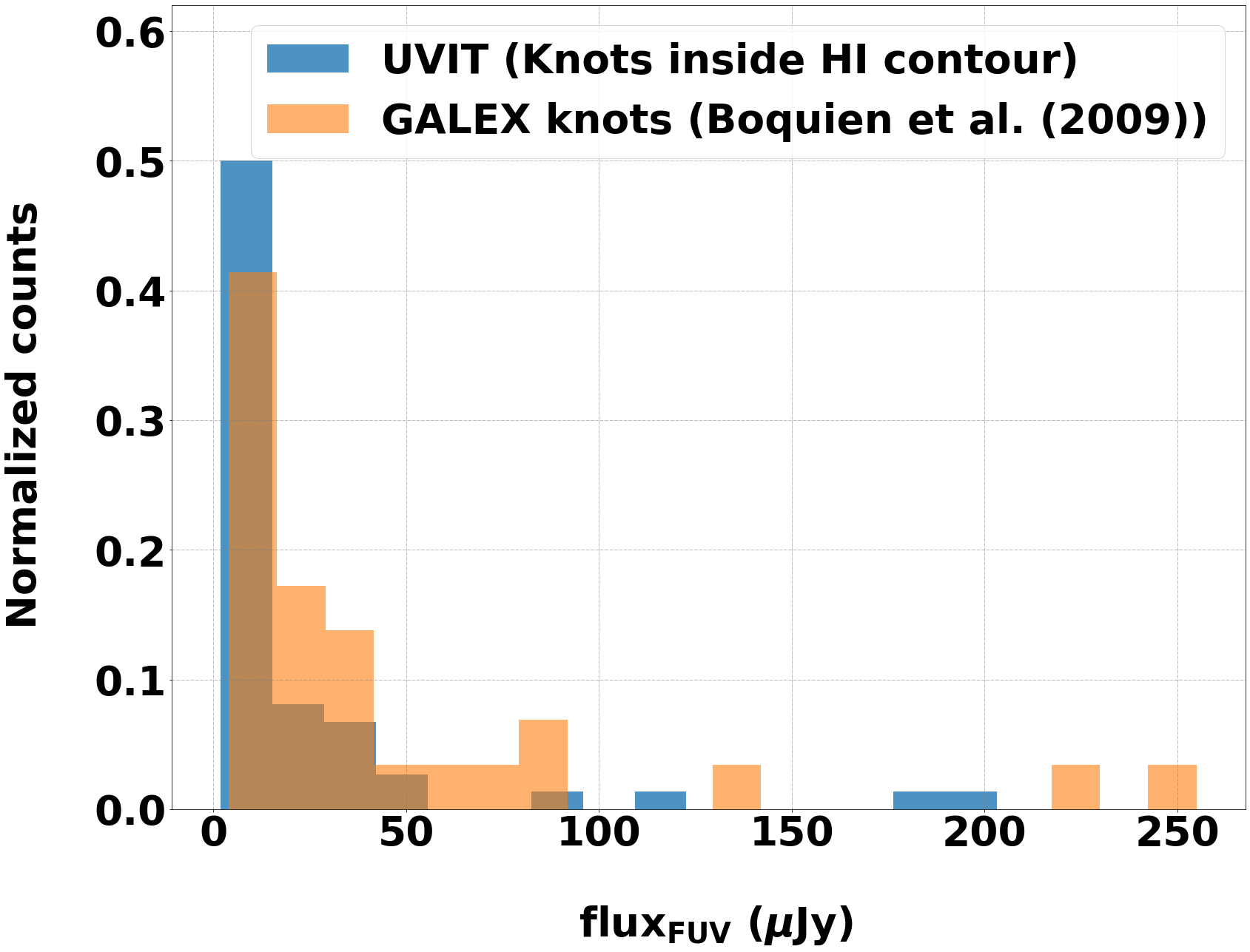}
\caption{Comparison of the flux values for the knots measured from UVIT and GALEX FUV imaging data. \citep{2009AJ....137.4561B}} 
\label{fig:uvit vs galex flux hist}
\end{figure}

Fig. \ref{fig:uvit vs galex flux hist} shows the distributions of fluxes (uncorrected for extinction) for the knots in UVIT and GALEX \citep{2009AJ....137.4561B}.

It is seen that the flux distribution for the knots in UVIT
has moved to the low flux values as compared to GALEX. This
can be attributed to the improved spatial resolution of UVIT in comparison with GALEX which enabled better deblending of structures.


\subsection{Slope of the UV continuum $\beta$ and extinction $A_{FUV}$}

The interstellar medium within the star forming knots can contain significant amount of dust. The UV radiation emitted by young massive O,B,A stars can get attenuated by dust. Dust grains can scatter and absorb UV radiation and this can greatly complicate the interpretation of the detected UV emission. Determining the level of
dust attenuation is  crucial in accurately deriving intrinsic UV luminosity and hence star formation rates. The slope of ultraviolet continuum has been proposed as a powerful diagnostic of dust attenuation in star-forming galaxies \citep{2012A&A...539A.145B,2011ApJ...726L...7O}. 
The UV continuum spectrum of star forming galaxies is characterised by the spectral index $\beta$ where $f_\lambda \propto \lambda^\beta$ \citep{1994ApJ...429..582C} for $\lambda >1200$ {\AA},  $f_\lambda$ (erg cm\textsuperscript{-2} s\textsuperscript{-1} {\AA}\textsuperscript{-1}) is the flux density
of the source. 
For the case of UVIT FUV and NUV passbands, 
\begin{equation}
   \hspace{1.5cm} \beta_{UVIT}=1.88(m_{FUV}-m_{NUV})-2.0
    \label{eq1}
\end{equation}  
where $m_{FUV}$ and $m_{NUV}$ are the magnitudes in FUV and NUV respectively.

\cite{1999ApJ...521...64M} (hereafter, M99) established a relationship  between the UV spectral slope $\beta$ and the ratio of far infrared (FIR) and UV fluxes for a sample of starburst
galaxies. This method relates the FIR and UV radiation
emitted from galaxies.  It is considered to be a powerful tool in recovering the UV radiation lost due to the dust, regardless of the geometry of the dust. 

We use the M99 relation for the starburst case to determine dust attenuation from $\beta$  which is given as,  

\begin{equation}
\hspace{1.5cm} A_{FUV}=4.43+1.99\;\beta
    \label{eq:AFUV}
\end{equation}
where $\beta$ is given by Eq. \ref{eq1} for our case. 

\begin{figure}
    \centering
    \includegraphics[scale=0.13]{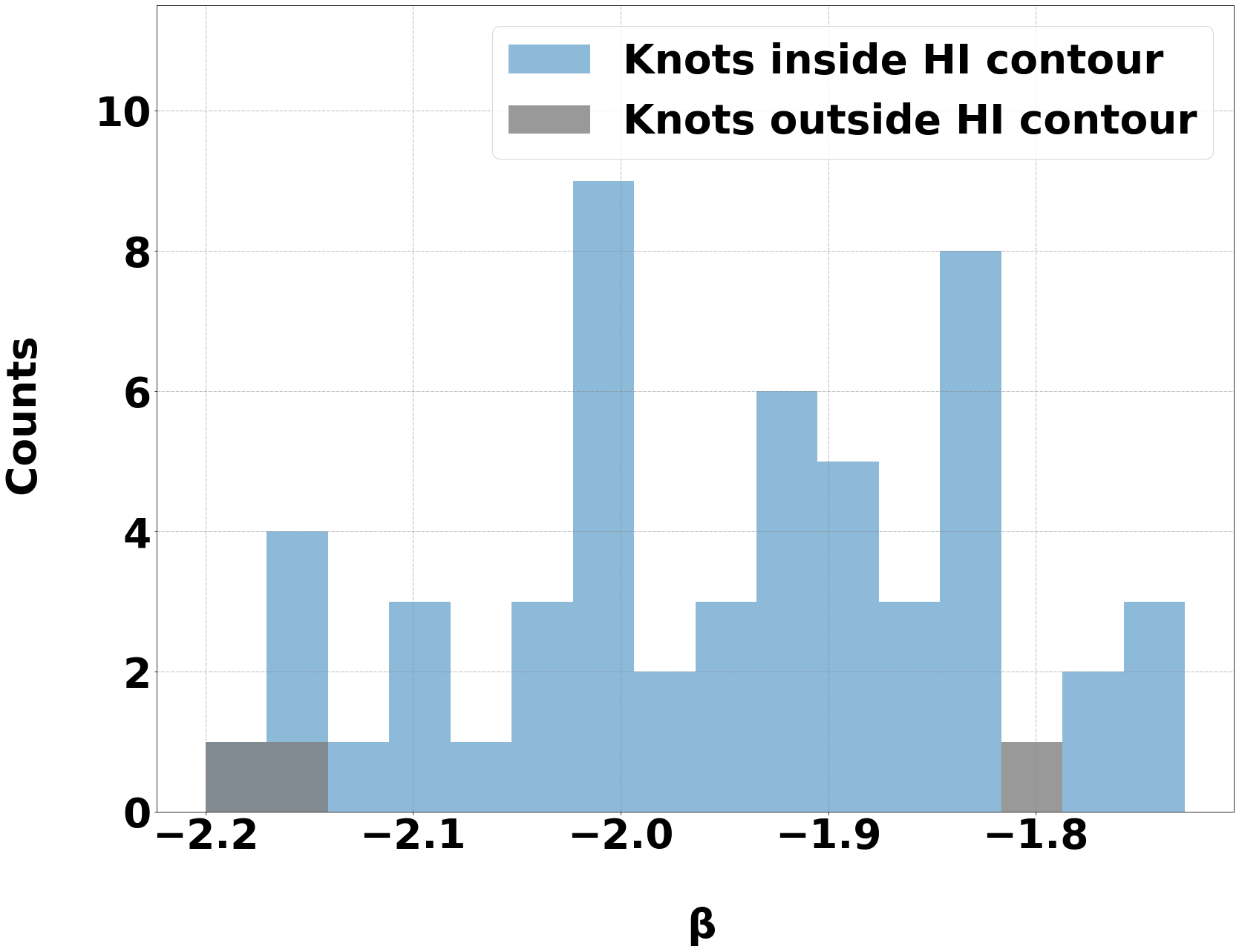}
    \centering \caption{Histogram of the values of the slope of the UV continuum $\beta$} 
\label{fig:histogram of beta}
\end{figure}

A histogram showing the slope of UV continuum $\beta$ for the selected knots in the NGC 5291 system is presented in Fig. \ref{fig:histogram of beta}. The numerical value of $\beta$ ranges from -2.18 to -1.73.

\begin{figure}
\centering
 \includegraphics[scale=0.13]{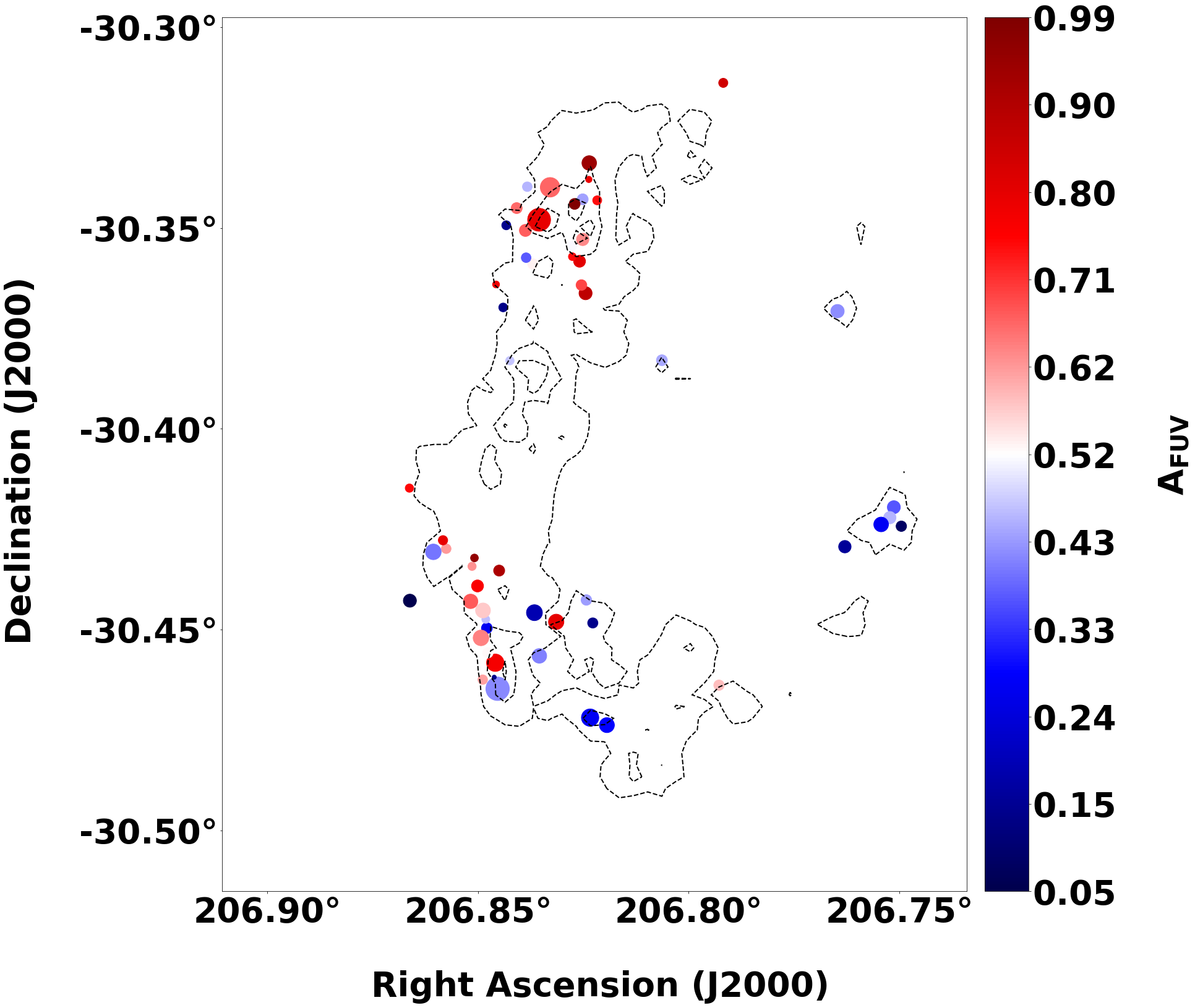}
 \caption{Spatial distribution of dust attenuation $A_{FUV}$ of the selected knots in the NGC 5291 system}
 \label{fig: extinction-map of 57 knots}
\end{figure}

Fig. \ref{fig: extinction-map of 57 knots} gives the spatial distribution of $A_{FUV}$ for the selected knots and the value of $A_{FUV}$ ranges from 0.05 to 0.99.

\subsection{Star formation rates of the knots} 

The measured FUV flux of the knots are corrected for extinction using the $A_{FUV}$ values computed for each knot. Ultraviolet flux is a direct tracer of ongoing star formation and the star formation rate (SFR) can be calculated from the extinction corrected UV luminosity \citep{2012ARA&A..50..531K}.

For the computation of the SFR, the following form of the relation is used which assumes a constant rate of star formation over a timescale of $10^8$ years, 
with a Salpeter initial mass function (IMF) \citep{1955ApJ...121..161S} from 0.1 to 100 M$_\odot$ as described in \cite{2006ApJS..164...38I} and in \cite{2008MNRAS.390.1282C}
\begin{equation}
SFR_{FUV}[M_\odot/yr] = \frac{L_{FUV}[erg/sec]}{3.83 \times 10^{33}}\times 10^{-9.51}
        \label{sfr}
\end{equation}
 where, $L_{FUV}$ is the extinction corrected FUV luminosity. 

 The total extinction corrected star formation rate, derived from the FUV flux, for the SF knots lying \textit{within} the HI contour (excluding that of NGC 5291 and the Seashell galaxies) amounts to  1.72 $\pm$ 0.04 M$_\odot$ yr$^{-1}$ and the same for the knots that lie \textit{outside} HI contour is 0.026 $\pm$ 0.004 M$_\odot$ yr$^{-1}$. The SFR of the galaxies, NGC 5291 and the Seashell is 1.93$\pm$ 0.21 M$_\odot$ yr$^{-1}$ and 1.16 $\pm$ 0.16 M$_\odot$ yr$^{-1}$ respectively. 

\section{Discussion}

There exists many observational studies on the  star formation and TDG formation in interacting systems in the UV using GALEX data
\citep{2005ApJ...619L..87H,2005ApJ...619L..91N,2009AJ....137.4643H,2009AJ....138.1911S,2009AJ....137.4561B}.
 \citet{2009AJ....137.4561B} made a detailed multi-wavelength (UV, infrared and H$\alpha$) analysis of six interacting systems with star forming regions. The interacting systems considered by them include NGC 5291, Arp 105, Arp 245, NGC 7252, Stephan's Quintet (SQ) and VCC 2062. \citet{2018A&A...614A.130G} performed a detailed study using UVIT on star formation in TDGs along the tails of the post-merger system NGC 7252. 
\begin{figure}
\centering
    \includegraphics[scale=0.13]{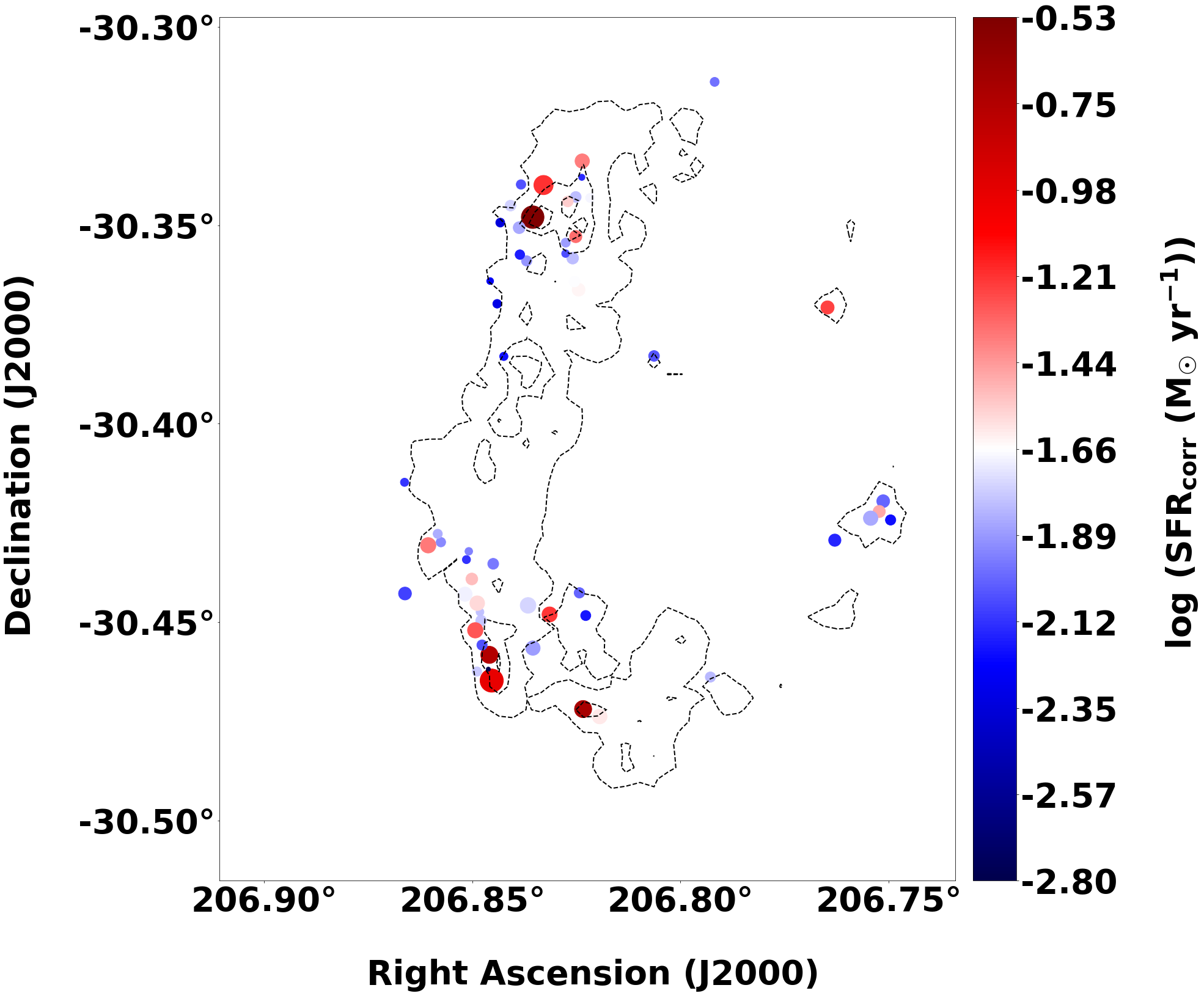}
    \caption{Spatial distribution of star formation rate SFR\textsubscript{FUV} of the selected knots in the NGC 5291 system}
    \label{fig:sfr of knots}
\end{figure}

The main concern in the estimation of SFRs using UV flux is the effect of dust attenuation. Dust plays a significant role in the attenuation of UV flux in galaxies.   
A common technique for measuring the extinction towards stars
in our Galaxy is to use the colour excess. If a star’s spectral type (and therefore its
intrinsic colour) is known, the extinction towards it can be determined. However, this technique cannot be  applied to galaxy systems, as the relative extinction at different wavelengths is sensitive to the unknown relative geometry of stars and dust, and differs for different
optical depths \citep{1998MNRAS.297..807T}.

The $L_{IR}/L_{UV}$ ratio has been identified as one of the powerful estimators of dust attenuation in star-forming galaxies e.g. \citep{2000ApJ...533..236G,2005ApJ...619L..51B}. 
From UV and FIR (Spitzer / Herschel) data available for NGC  5291, one could directly measure the dust attenuation from FIR/UV, but not at the spatial resolution of UVIT.
The relation established by \citet{1999ApJ...521...64M} between the slope of the rest frame
UV continuum $\beta$ and dust attenuation deduced from their $L_{IR}/L_{UV} - \beta$ relation for local
starburst galaxies is hence used
to determine the extinction towards the knots.
 
\begin{figure}
\centering
    \includegraphics[scale=0.13]{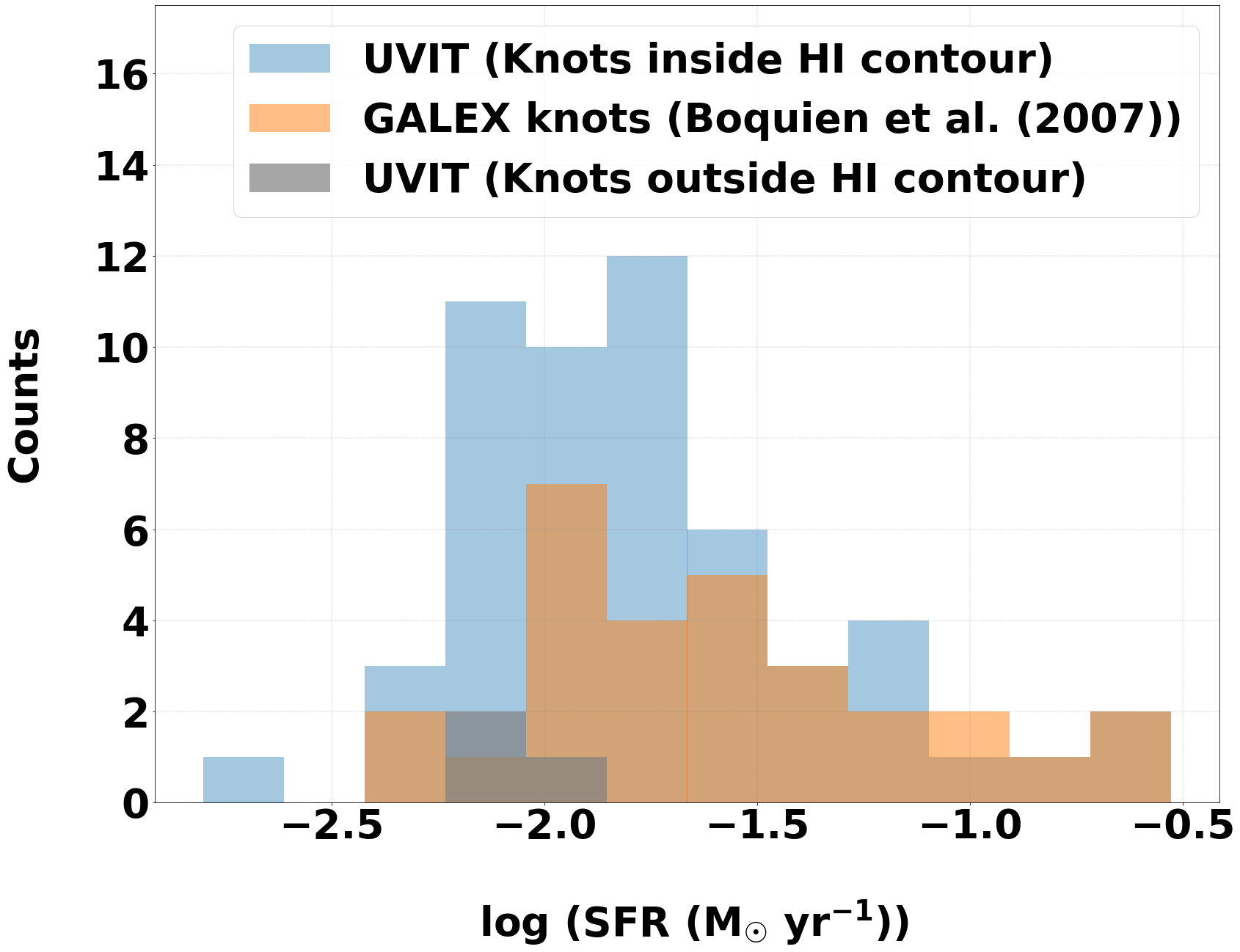}
    \caption{Histograms of SFRs (corrected for extinction) of the knots in the current  study and SFRs of knots given in \citet{2007A&A...467...93B} for the NGC 5291 interacting system}
    \label{fig:uvit galex sfr comp}
\end{figure}

In this paper, we have analysed the extinction in the star forming regions associated with the NGC 5291 interacting system 
using high resolution UVIT data. 
The SFR for the knots of the selected knots of the NGC 5291 system has been computed from corrected UV luminosities. The spatial distribution of extinction corrected FUV star formation rate (SFR\textsubscript{FUV}) of the selected knots in NGC 5291 interacting system is presented in Fig. \ref{fig:sfr of knots}. The extinction corrected SFR (log SFR) of the knots ranges from -2.80 to -0.53.
The M99 relation provides us with accurate estimates of the attenuation for starburst galaxies \citep{2012A&A...539A.145B,2011ApJ...726L...7O}.
Considering the ongoing star formation in the NGC 5291 system to be more like a starburst, the M99 relation is used here for the estimation of extinction.
 This is the first time such an analysis is performed on the NGC 5291 system. 
  UVIT has resolved  the star forming regions in comparison to the previous UV mission GALEX. Several of the knots that appeared as a single entity in the GALEX images have been resolved into two or more knots in the UVIT images. This enabled the estimation of the extinction to the smallest scales that have been ever possible for NGC 5291.
   The UVIT based extinction corrected SFRs for the selected knots in the current study is compared with that of the previously measured SFR values given in \citet{2007A&A...467...93B} and shown in Fig. \ref{fig:uvit galex sfr comp}. 

\begin{figure}
\centering
    \includegraphics[scale=0.13]{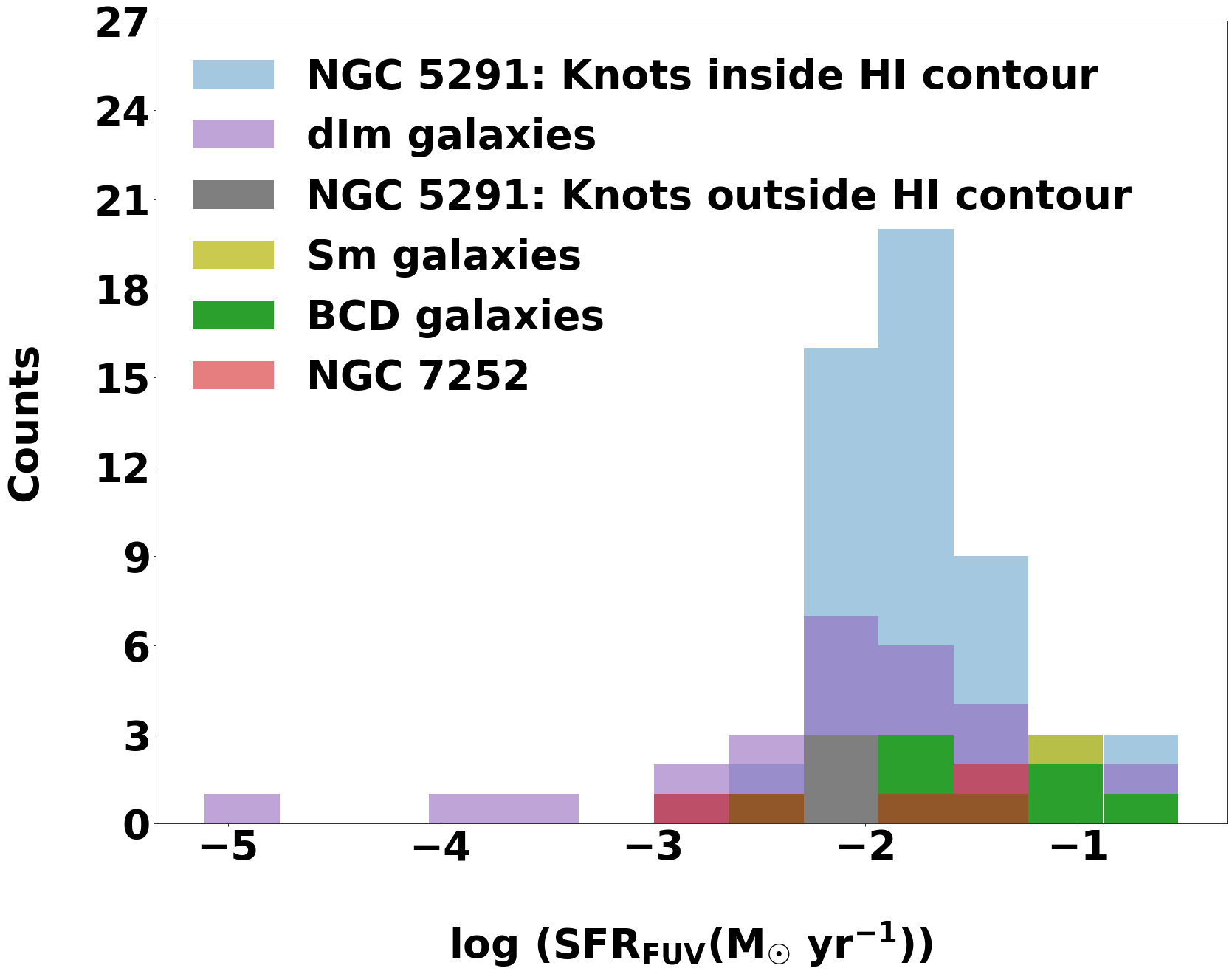}
    \caption{Histograms showing the SFR values of the selected knots of the NGC 5291 system as per current study, SFRs of BCD (8 samples), Sm (7 samples) and dIm (29 samples) galaxies in the nearby universe from \citet{2010AJ....139..447H} and uncorrected SFR of knots of the NGC 7252 system given in \citet{2018A&A...614A.130G}.}
    \label{fig:histogram of sfr}
\end{figure}

In order to check whether the recently formed TDGs exhibit SFR values comparable to other dwarf galaxies of similar stellar masses in the local universe, we compared the extinction corrected SFR values of the selected knots in the NGC 5291 system with the SFR values of  dwarf galaxies (eg. BCD), Sm and dIm galaxies in the local universe  as determined using GALEX FUV data by \citet{2010AJ....139..447H} as well as with the uncorrected SFR of knots in NGC 7252 system as given in \citet{2018A&A...614A.130G}.
Histograms for the SFR values of the aforementioned systems are shown in Fig. \ref{fig:histogram of sfr}. We note that the values are very similar to the uncorrected SFR of star forming knots detected in NGC 7252 from UVIT imaging.
From the histogram, it is further noted that log SFRs of the selected knots in the NGC 5291 system is greater than -3. The distribution of SFRs in the knots is similar to those of dIm galaxies for log SFR greater than -3. The maximum value of log SFR is similar for the knots in the NGC 5291 system, dIm and BCD galaxies. It is also seen that many of the knots associated with the NGC 5291 system have high SFRs similar to BCD galaxies; this is characteristic of Category 1 TDGs. 
The three known tidal dwarf galaxies in the system, located to the north - NGC 5291N, south - NGC 5291S \citep{1998A&A...333..813D,1999IAUS..186...61D,2006ApJ...640..768H} and south-west - NGC 5291SW have SFR values of 0.30 M$_{\odot}$ yr$^{-1}$, 0.30 M$_{\odot}$ yr$^{-1}$ and 0.22 M$_{\odot}$ yr$^{-1}$ respectively. They are located far from the center of the system and are located at peaks in the HI column density map. The highest log SFR values reported by \citet{2010AJ....139..447H} for
dIm, BCD and Sm galaxies in their samples are -0.62, -0.62 and
-0.94 respectively, while the log SFR of the three known TDGs
in the NGC 5291 system are -0.52 (NGC 5291N), -0.52 (NGC 5291S)
and -0.66 (NGC 5291SW).

\section{SUMMARY}
The star forming knots in the NGC 5291 interacting system, which
includes three bonafide TDGs and several TDG candidates was investigated using high-resolution FUV and NUV data from AstroSat’s
UVIT.  
The star-formation activity in the selected star-forming knots was further studied by determining their
extinction-corrected SFR values. 
The results are summarized as below:
\begin{itemize}
    \item[-] A total of 57 star-forming knots have been identified as being part of the NGC 5291 interacting system.
    \item[-] The resolved star-forming knots range in size from 1.4 kpc to 11.4 kpc.
    \item[-] In comparison to the previous UV imaging at lower resolution, we have 12 new detections. The higher resolution of UVIT has allowed for better de-blending of the structures. Several of the knots in the NGC 5291 system which appeared as single star forming regions in GALEX images are well resolved into smaller star forming knots in the UVIT images.
    \item[-] The extinction towards each of the resolved star-forming knots and the main body of the NGC 5291 interacting system was computed using the slope of the UV continuum and hence the extinction-corrected SFR was determined. The total extinction-corrected SFR of the knots (inside and outside HI contour), excluding NGC 5291 and Seashell galaxies, is estimated as 1.75 $\pm$ 0.04 M$_{\odot}$ yr$^{-1}$.
    \item[-] Comparison of the NGC 5291 system with NGC 7252 using UVIT data showed that the SFR for both the systems are similar. Also the comparison with independent dwarf galaxy populations (BCD, Sm and dIm galaxies) in the nearby Universe showed that many of the knots in the NGC 5291 system have SFR values comparable to the SFR of BCD galaxies.
\end{itemize}

\section*{Acknowledgements}
The authors RR and GS acknowledge the financial support of ISRO under
AstroSat archival Data utilization program (No. DS\_2B-13013(2)/9/2020-Sec.2).  This publication uses data from the AstroSat mission of the Indian Space Research Organisation (ISRO), archived at the Indian Space Science Data Centre (ISSDC). RR acknowledges visiting associateship of IUCAA, Pune. KG and LC acknowledge support from the Australia-India Council/Department of Foreign Affairs and Trade (via grant AIC2018-067). We thank Aaron Robotham for help with ProFound. SS acknowledges support from the Science and Engineering Research Board of India through Ramanujan Fellowship and POWER grant (SPG/2021/002672). LC acknowledges support from the Australian Research Council Discovery Project and Future Fellowship funding schemes (DP210100337, FT180100066). This work used Astropy, Matplotlib and Reproject software packages \citep{astropy:2013, astropy:2018, astropy:2022, Hunter:2007, robitaille_thomas}.
\section*{Data Availability}

The Astrosat UVIT imaging data underlying this article are available in ISSDC Astrobrowse archive (https://astrobrowse.issdc.gov.in/astro\_archive/archive/Home.jsp) and can be accessed with proposal ID: G07\_003.

\newpage
\bibliographystyle{mnras}
\bibliography{References}
\end{document}